# ConocoPhillips' share price model revisited
## Ivan Kitov

Institute for the Geospheres' Dynamics, Russian Academy of Sciences

**Abstract**
Three years ago we found a statistically reliable link between ConocoPhillips' (NYSE: COP) stock price and the difference between the core and headline CPI in the United States. In this article, the original relationship is revisited with new data available since 2009. The agreement between the observed monthly closing price (adjusted for dividends and splits) and that predicted from the CPI difference is confirmed. The original quantitative link is validated. In order to improve the accuracy of the COP price prediction a series of advanced models is developed. The original set of two major CPIs is extended by smaller components of the headline CPIs (e.g. the CPIs of motor fuel and housing energy) and several PPIs (e.g. the PPIs of crude oil and coal) which may be inherently related to ConocoPhillips and other energy companies. These advanced models have demonstrated much lower modeling errors with better statistical properties. The earlier reported quasi-linear trend in the CPI difference is also revisited. This trend allows for an accurate prediction of the COP prices at a five to ten year horizon.

**Introduction**
Three years ago we introduced a model based on the link between consumer and stock prices [1]. We found a statistically reliable relationship between ConocoPhillips' stock price and the difference between the core and headline CPI in the United States. A similar relationship was also estimated for Exxon Mobil. It is instructive to revisit the original quantitative relationships with the relevant data available since 2009 in order to estimate them qualitatively and statistically. In this article, we focus on the evolution of ConocoPhillips' share price.

Originally, the agreement between the observed monthly closing price (adjusted for dividends and splits) and that predicted from the (seasonally not adjusted) CPI difference was relatively good and our tentative model covered the period between 1982 and 2009, which was split into two segments in order to match the turn in the trend observed in the CPI difference between 1998 and 2002. The initial model based on two major CPIs was rather crude, however, and did not exercise numerous options associated with smaller CPI components directly connected to energy prices. We have investigated the whole S&P 500 list since 2009 and found hundreds of statistically robust quantitative models based on finer consumer price indices [2, 3]. It would not be an exaggeration to conclude that the original approach has been significantly improved and the advanced models have shown a much higher predictive power, reliability and accuracy. Therefore, it is mandatory to apply the advanced approach to COP's stocks.

Our pricing model assumes that the future of selected stocks is not unpredictable. Despite the bias of market participants, who are definitely convinced that all available information is already priced in, we have found many companies with stocks described by deterministic models based on various CPIs. This unaccounted information allows outperforming the market and its existence does not contradict common wisdom and scientific knowledge. There are true links between measured variables which we do not know yet. Accordingly, there should exist many market features and processes currently inaccessible, but fully objective and describing the evolution of prices far beyond the contemporary market paradigm. Currently, there are many models and a huge number of tools related to stock pricing. Our pricing concept shows that these models and tools are likely of a limited usage only because they are constrained by the convention of price stochasticity and unpredictability. No of these ideas or techniques are borrowed and thus we omit usual review of the literature devoted to stock markets as irrelevant.

The remainder of this paper is organized as follows. Section 1 presents linear trends in the difference between the core and headline CPI observed in the past and predicts the evolution at a several year horizon. In Section 2, the evolution of COP price is predicted as a linear function of the CPI difference. The newly available estimates of the stock price and defining CPIs allow validation of the original model. Section 3 extends the set of defining CPIs and introduces PPIs as the COP price drivers. A linear time term and an intercept are added to the model. All these improvements allow a more accurate description with the best model having a standard error of $3.96 between July 2003 and March 2012.

## 1. The concept and data

We have revealed many sustainable (quasi-) linear trends in the differences between consumer and producer price indices [4-6]. At first, it was found that the difference between the core CPI, *CC*, and the headline CPI, *C,* can be approximated by a linear time function:

$$dCPI(t) = CC(t) - C(t) = A + Bt \qquad (1)$$

where *dCPI(t)* is the difference, *A* and *B* are empirical constants, and *t* is the elapsed time. Therefore, the distance between the core CPI and the headline CPI is a linear function of time, with a positive or negative slope *B*. Figure 1 displays this difference from 1960 to 2012. Both variables are not seasonally adjusted and are borrowed from the Bureau of Labor Statistics [7]. There are three distinct periods of linear time dependence: from 1960 to 1980, from 1980 to 1998, and from 2002 to 2008. The second period is characterized by a linear trend with *B*=+0.65, and the third one has a larger negative slope of *B*=-1.52.

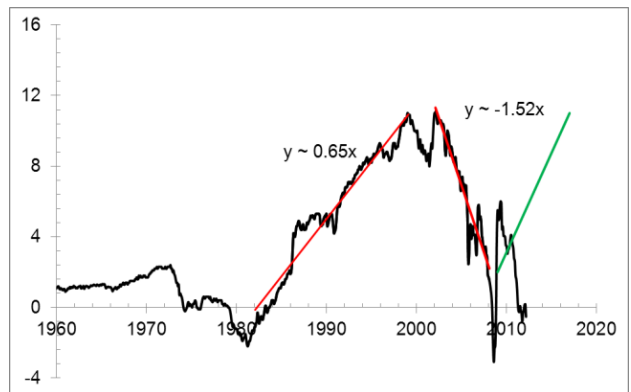

**Figure 1. The difference between the core and headline CPI as a function of time. One can distinguish three periods of quasi-linear behavior with two distinct turning points. For the second and third periods, linear regression lines are characterized by slopes *B*=+0.65 and *B*=-1.52, respectively. Solid green line represents the trend between 2009 and 2015 predicted as a mirror reflection of the previous trend, i.e. a straight line with *B*=+1.52.**

Accordingly, there are two turning points or short transition intervals - between 1980 and 1981 and from 1999 to 2002. Since 2009, the difference has been passing the third turning point accompanied by a high degree of volatility. Similar behavior was observed between 1999 and 2002. In the past, the linear trends were very strong attractors to all deviations. Therefore, it is likely that in the near future a new linear trend will emerge, which may repeat the previously observed duration and slope. In Figure 1, green solid line represents the trend between 2009 and 2015 predicted as a mirror reflection of the previous trend between 2002 and 2008. Basically, the difference has to grow from 1 unit of index in 2009 to 11 units in 2015. In the beginning of 2012, the actual value was negative but close to zero.

Our approach to stock pricing is almost trivial. Imagine that one has to predict (describe) the evolution a share price for an energy company. It would not be a big mistake to assume that this share price is likely to be driven by the change in the overall energy price, which can be expressed by the price index including all energy prices (e.g. the headline CPI). Alternatively, some components of the overall energy category might be in play. Even if the studied company does not change its production the overall increase in the price of its product (and thus some consumer prices) should be manifested in its profit and higher (or smaller) share price. On the other hand, when other prices (e.g. the core CPI) rise faster than the energy price index (say, 10% *vs.* 1% per year, respectively) one should not expect the energy company to gain extra pricing power. The company would rather suffer a share price decline. Thus, considering the secular increase in the overall price level, it is not the absolute change in energy prices what affects the stock price but its current deviation from some energy independent price.

Without loss of generality, we have proposed to use the simplest model as based on the difference between the core and headline CPIs. No time lags between these indices were introduced in the beginning. (However, we allowed the stock price to lead the CPI difference.) We also ignored the sensitivity of the share price to the change in the core and headline CPIs and used them with the same weight of 1.0. When two defining CPIs evolve at quite different rates, one has to apply weighting, i.e. to introduce independent coefficients to both defining CPIs, in order to equalize their respective inputs.

The headline CPI includes all kinds of energy and thus provides the broadest proxy to the energy price index. The core CPI excludes energy (and food) and thus may represent the energy independent and dynamic reference. In the initial approach, we assumed the presence of a linear link between a stock price (*COP*) and the difference between the core and headline CPI:

$$COP(t) = a + bdCPI(t + T) \qquad (2)$$

where *a* and *b* are empirical constants; *T* is the time delay between the stock and the CPI change, the CPI may lead or lag the price. Constants in (2) should be determined for each linear trend period separately. This implies the possibility of structural breaks in relationship (2) due to the turn to a new trend. One may suggest that any new trend manifests some deep structural changes in the overall economic behavior. Otherwise, there would be no change in the trends.

**2. COP model revisited**

Three years ago, the evolution of a ConocoPhillips' stock price was modeled as a linear function of the *dCPI*. Because we tested the general approach, only the trial-and-error method was applied and we sought for the overall visual fit between the observed and predicted prices, the latter is a scaled *dCPI(t)*. In the original model, the best fit coefficients between 1998 and 2009 were as follows:

$$COP(t) = -6.0dCPI(t+2) + 80 \qquad (3)$$

The time lag of 2 months better fits the price fall in 2008. The slope was -6.0, i.e. the dCPI change by 1.0 has to be mapped into the price fall of $6. The intercept $80 implies that the long term level of COP price is $80 when *dCPI*=0.

We have amended the original model and re-estimated all coefficients for the period between 1998 and 2012 using the LSQ method. The new model is as follows:

$$COP(t) = -5.35dCPI(t+1) + 72.3; \sigma=\$7.87 \qquad (4)$$

where $\sigma$ is the standard model error for the studied period. Relationship (4) is different from (3) and provides a slightly better overall fit. Figure 2 illustrates the predictive power of the model.

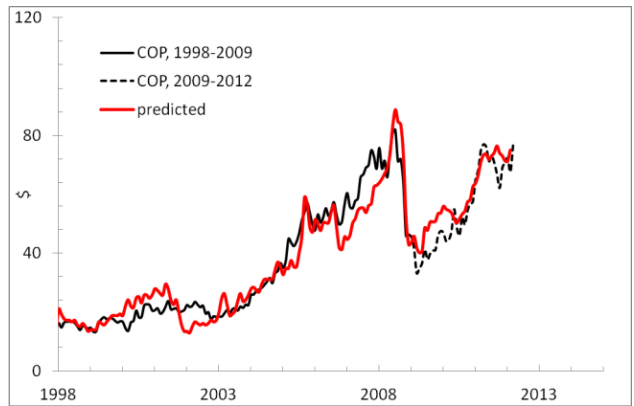

**Figure 2. The predicted COP share price (red line), i.e. the scaled difference between the core CPI and the headline CPI from 1998 to 2012. Between 1998 and 2009 the historical prices for COP are shown by solid black line, and the period since 2009 with new data is shown by dashed line.**

When extrapolated in the past, i.e. before 1998, relationship (4) fails to predict the price as red line in Figure 3 reveals. Judging from the discrepancy before 1998, one might wrongly suggest that the *dCPI* has no predictive power. Let's return to Figure 1, however, which shows that the linear trend before 1998 was positive and after 2002 – negative. Econometrically speaking, there was a structural break in the difference between the core and headline CPI. Hence, it would not be a big mistake to suggest that some inherent directions of pricing powers swopped between 1998 and 2002. It is reasonable to assume that the sign of slope in (4) before 1998 was opposite to that from 2002 to 2008. After reversing the sign and calibrating relevant amplitude and level between 1987 and 1998 we have obtained a much better fit, as shown by green line in Figure 3, using the following function:

$$COP(t) = 2.2dCPI(t+1) - 7; 1987<t<2002 \qquad (5)$$

Finally, a complete COP price prediction between 1987 and 2009 is obtained. Before 1987, the predicted curve in Figure 3 sinks below the zero line. There is no special need to describe the price in the early 1980s using the CPI difference. As shown in [4, 5], all subcategories of the consumer price index, except the index of energy, are parallel before 1982. Therefore, the difference between any two indices, including the headline and core CPI, is constant, i.e. it contains no information on the changes in stock prices. This was the result of the CPI measuring procedures. New definitions and procedures were introduced between 1977 and 1982; they gave birth to numerous independently evolving CPIs.

Three years ago we suggested the COP price to follow the new trend in the *dCPI* (green line) in Figure 1 with possible change in the slope sign. Figure 2 definitely shows that there has been no change in the sign since 2009 and likely no new trend has emerged. The turn did not happen yet. However, we are waiting for a turn to the new trend when oil price will go down. On the other hand, the original model still works well and predicts larger movements in the price. Overall, our initial pricing model has matched the challenge of new data. The difference between the core and headline CPI gives a good approximation to the evolution of COP price.

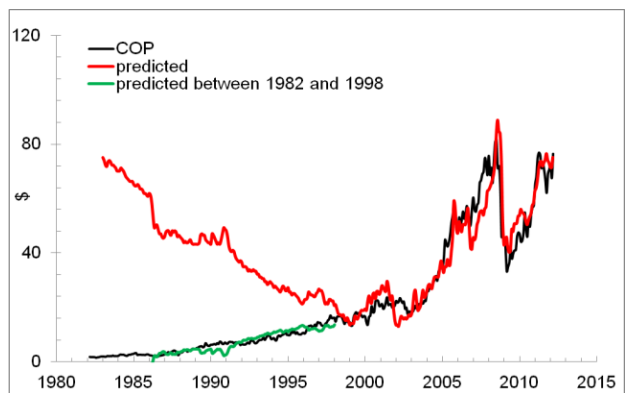

**Figure 3. The observed and predicted COP price.**

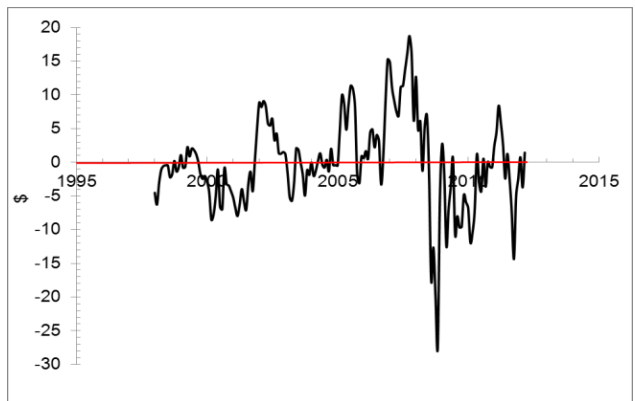

**Figure 4. The difference between the observed and predicted time series.**

Figure 4 depicts the model error since 1998. There were periods of large deviations which all ended on the predicted curve. This effect allows formulating a hypothesis that any current deviation will return to the predicted curve which one may consider as a fundamental one. The most recent estimates show that the observed price has returned to its fundamental level after a short period of undervaluation. Potentially, this is a short-term investment idea. The original model is crude, however, and we are looking for a better description with finer CPIs and PPIs.

## 3. Advanced models

The original model does not match high standards of quantitative modeling and statistical assessment. Both CPIs depend on many other goods and services, what introduces high measurement noise in the model. Also, both CPIs have the same weight (1.0) and cannot lead or lag behind the modeled price or each other. Apparently, it can be some non-zero lag between the change in energy price and in prices of energy companies. Therefore, we have extended the model and described the evolution of a share price as a weighted sum of two individual consumer price indices (or PPIs) selected from a large set of CPIs. We allow both defining CPIs (PPIs) lead the modeled share price or lag behind it. Additionally, we introduced a linear time trend on top of the earlier introduced intercept. In its general form, the pricing model is as follows:

$$sp(t_j) = \Sigma b_i \cdot CPI_i(t_j - \tau_i) + c \cdot (t_j - 2000) + d + e_j \qquad (6)$$

where $sp(t_j)$ is the share price at discrete (calendar) times $t_j$, $j=1,\ldots,J$; $CPI_i(t_j - \tau_i)$ is the $i$-th component of the CPI with the time lag $\tau_i$, $i=1,\ldots,I$; $b_i$, $c$ and $d$ are empirical coefficients of the linear and constant term; $e_j$ is the residual error, which statistical properties have to be scrutinized.

By definition, the bets-fit model minimizes the RMS residual error. The time lags are expected because of the delay between the change in one price (stock or goods and services) and the reaction of the other prices. It is a fundamental feature of the model that the lags in (6) may be both negative and positive. In this study, we limit the largest lag to thirteen months. Apparently, this is an artificial limitation and might be changed in a more elaborated model. In any case, a thirteen-month lag seems to be long enough for any price signal to pass through.

System (6) contains $J$ equations for $I+2$ coefficients. Since the sustainable trends last more than nine years, the share price time series has more than 100 points. Due to the negative effect of a larger set of defining CPI components we limit the dimension to ($I=$) 2 in all models. To resolve the system, standard methods of matrix inversion are used. A model is considered as a reliable one when the defining CPIs are the same during the previous eight months. The number and diversity of CPI subcategories is a crucial parameter. In this study we progressively extend the set of defining components

So far, we have tested one principal pair of CPIs: C and CC. Now we try two more pairs: CC and the index of energy, E, as well as the pair the PPI and the producer price index of crude oil, OIL. The best fit model is obtained with the pair PPI and OIL ($\sigma=\$5.98$):

$$COP(t) = 3.740C(t) - 5.148CC(t-12) + 6.73(t-2000) + 223.48; \; \sigma=\$6.21 \qquad (7)$$
$$COP(t) = -2.774CC(t-12) + 0.460E(t) + 10.60(t-2000) + 345.89; \; \sigma=\$5.98 \qquad (8)$$
$$COP(t) = 1.598PPI(t-0) + 0.029OIL(t-2) - 6.43(t-2000) - 102.89; \; \sigma=\$6.35 \qquad (9)$$

Figures 5 through 7 depict the observed and predicted monthly prices from (7) through (9). In Figure 6, the advanced model based on the CPI of energy accurately predicts the COP price. Lately, there was one major (negative) deviation from the predicted price - in the fourth quarter of 2011. It ended on the fundamental price curve in January 2012. An investor could use the knowledge of the transient character of such a deviation and foresee the future return. Moreover, any large deviation likely gives a good investment idea.

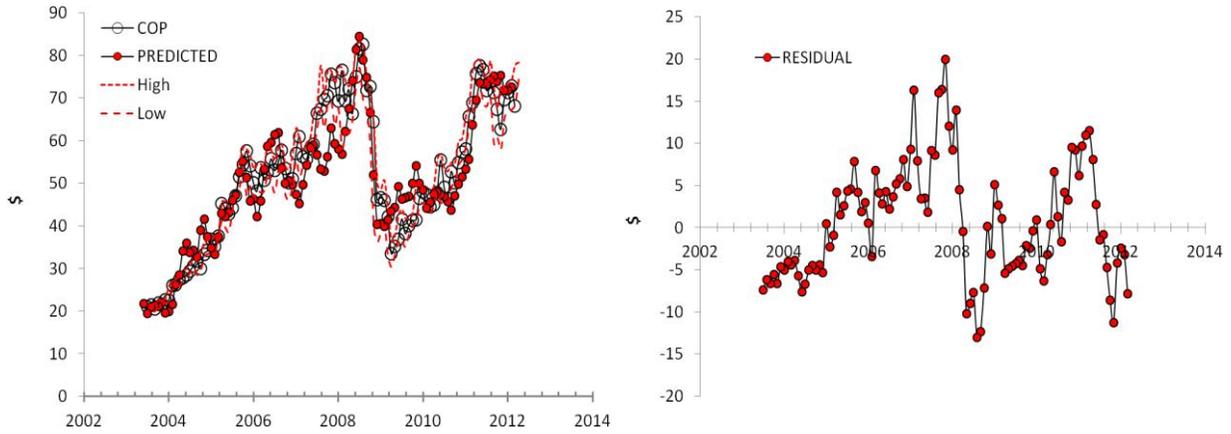

**Figure 5.** Left panel: The observed (monthly closing) COP price and that predicted by relationship (7) from the headline and core CPI. The high and low monthly prices provide an estimate of uncertainty. Right panel: The model residual.

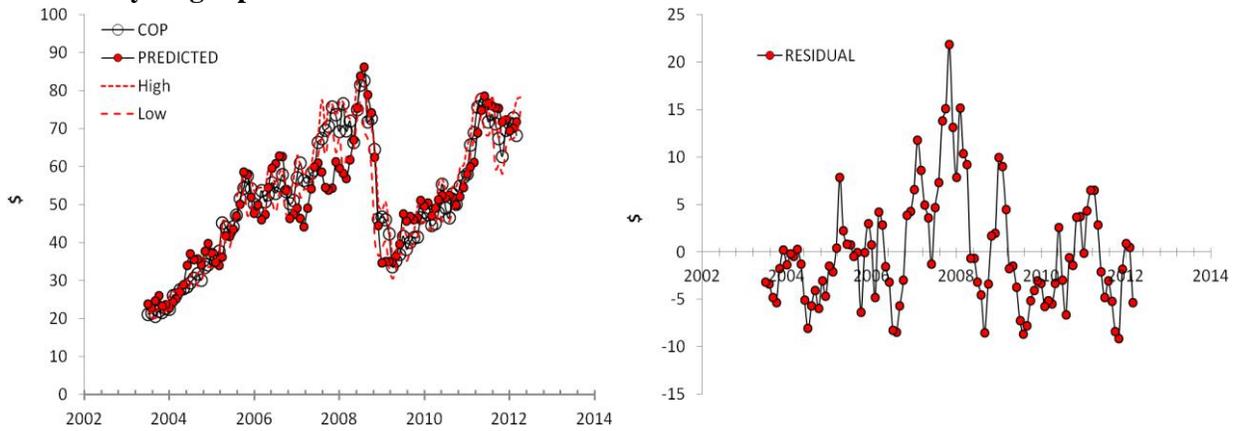

**Figure 6.** Left panel: The observed (monthly closing) COP price and that predicted by relationship (8) from the core and energy CPI. The high and low monthly prices provide an estimate of uncertainty. Right panel: The model residual.

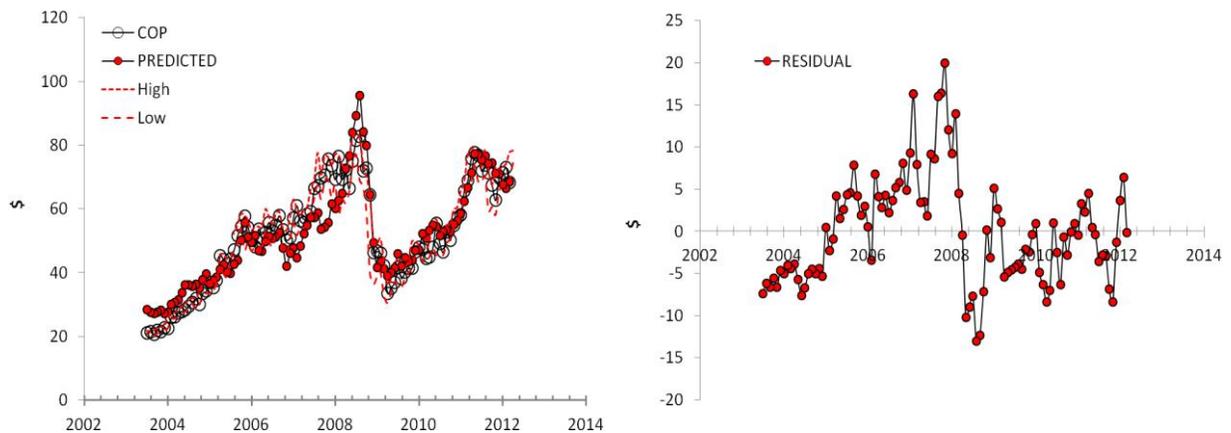

**Figure 7.** Left panel: The observed (monthly closing) COP price and that predicted by relationship (9) from the overall and oil PPI. The high and low monthly prices provide an estimate of uncertainty. Right panel: The model residual.

The next obvious step is to use a wider range of CPI and PPI components. Keeping in mind the inherent relation of ConocoPhillips to energy, we have selected the following indices:

C -        the headline CPI;
F -        the consumer price index of food and beverages;
H -        the consumer price index of housing;
FU -       the consumer price index of fuels and utilities (part of H);
HHE -      the consumer price index of household energy;
CE -       the headline CPI less energy;
CC -       the core CPI;
E -        the consumer price index of energy;
MF -       the consumer price index of motor fuel;
GAS –      the producer price index of natural gas;
COAL –     the producer price index of coal;
EL –       the producer price index of electricity;
OIL -      the producer price index of crude petroleum (dimestic production);
PPI -      the overall PPI

All pairs of these CPIs and PPIs were used as defining parameters and the best model was obtained with the overall PPI and PPI of coal:

$$COP(t) = -0.615 COAL(t-0) + 1.269 PPI(t-0) + 4.02(t-2000) - 105.35; \; \sigma = \$3.96 \qquad (10)$$

Figure 8 depicts the model and its error between 2003 and 2012. The standard error is now only $3.96, i.e. 67% of that in model (8). This is a tremendous improvement, which is successfully accompanied by a smoother distribution of the error over time. The price index of coal influences the COP price negatively: increasing coal price suppresses COP stocks.

Model (10) is the best among all studied models and has been also the best during the previous 8 months (see Table 1). In other words, the overall and coal PPIs give the smallest RMS residual since August 2011. It was likely the best model before 2011.

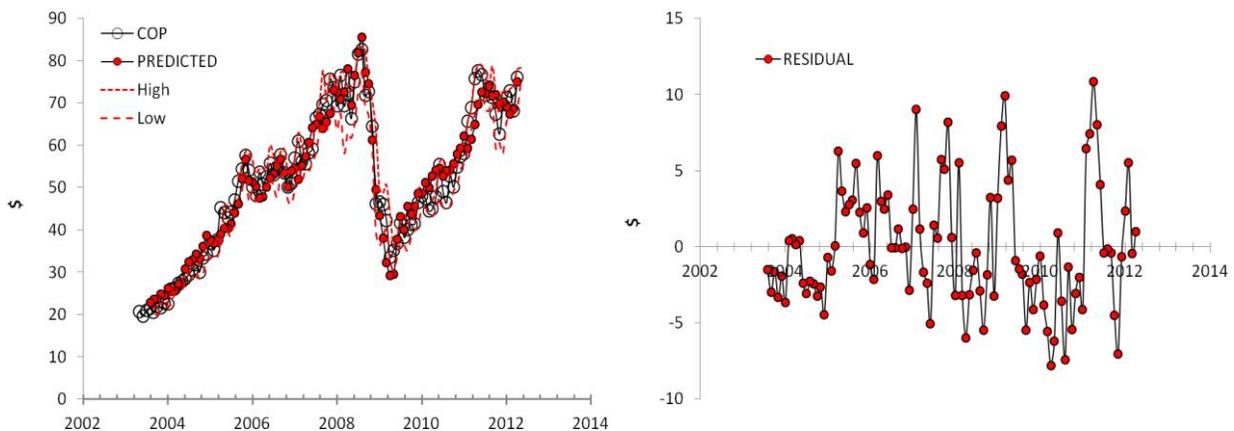

**Figure 8. Left panel: The observed (monthly closing) COP price and that predicted by relationship (10) from the overall and coal PPI. The high and low monthly prices provide an estimate of uncertainty. Right panel: The model residual.**

**Table 1. The defining PPIs, lags and coefficients of the best fit models since August 2011.**

| Month | $CPI_1$ | $Lag_1$ | $b_1$ | $CPI_2$ | $Lag_2$ | $b_2$ | $c$ | $d$ |
|---|---|---|---|---|---|---|---|---|
| March | COAL | 1 | -0.615 | PPI | 1 | 1.2687 | 4.023 | -105.35 |
| February | COAL | 1 | -0.6145 | PPI | 1 | 1.2685 | 4.01 | -105.313 |
| January, 2012 | COAL | 1 | -0.6145 | PPI | 1 | 1.2685 | 4.0133 | -105.325 |
| December | COAL | 1 | -0.6153 | PPI | 1 | 1.2692 | 3.9838 | -105.146 |
| November | COAL | 1 | -0.6147 | PPI | 1 | 1.2697 | 3.9549 | -105.119 |
| October | COAL | 1 | -0.6147 | PPI | 1 | 1.2698 | 3.9551 | -105.157 |
| September | COAL | 1 | -0.614 | PPI | 1 | 1.2713 | 3.9831 | -105.647 |
| August, 2011 | COAL | 1 | -0.6113 | PPI | 1 | 1.2755 | 3.9479 | -106.445 |

**Conclusion**

The initial COP price model is validated by the estimates of the headline and core CPI available since 2009. The model has accurately predicted the bottom price of the 2008 fall and the following recovery to the current level. However, the initial model does not provide (and did not aim to provide) the best statistical description of the COP price evolution.

A series of advanced price models has been developed as based on an extended set of CPIs and introduction of producer price indices as COP price drivers. The best model with the PPI of coal has reduced the original standard error by a factor of 2. This is a dramatic improvement accompanied by a very high stability of the best model, i.e. defining PPIs, their lags and coefficients, over time.

One may extend the set of defining indices and probably find a better model. We suggest that such an exercise would need some new CPIs or PPIs to be compiled, which are not officially published by the BLS. These CPIs/PPIs should include goods and services most influential for ConocoPhillips and exclude those introducing noise in the time series.